\documentclass[epj]{svjour}
\usepackage{graphics}
\usepackage{epsfig}
\begin{document}

\title{Application of the Extended Pairing Model to Heavy Isotopes}

\author{V. G. Gueorguiev\thanks{Conference presenter}\inst{,1 \and 2} \and
Feng Pan\inst{3} \and J. P. Draayer\inst{2}}

\institute{Lawrence Livermore National Laboratory, Livermore, CA 
94550, USA \and
Department of Physics and Astronomy, Louisiana State University, 
Baton Rouge, LA 70803, USA \and Department of Physics, Liaoning 
Normal University,
Dalian, 116029, P. R. China}

\date{\today}

\date{Received: October 21, 2004 / Revised version:  November 11, 2004}

\abstract{Relative binding energies (RBEs) within three isotopic chains ($^{100-130}$Sn,
$^{152-181}$Yb, and $^{181-202 }$Pb) have been studied using the exactly solvable 
extended pairing model (EPM) \cite{FPan04}. The unique pairing strength $G$, which 
reproduces the experimental RBEs, has been determined. Within EPM,  $\log(G)$ is 
a smooth function of the model space dimension $\dim(A)$, 
as expected for an effective coupling strength. In particular, for the Pb and Sn isotopes 
$G$ can be described by a two parameter expression that is inversely proportional to 
the dimensionality of the model space, $G=\alpha\dim(A)^{-\beta}$ with $\beta\approx 1$.
\PACS{
{21.10.Dr}{Binding energies}  \and
{71.10.Li}{Pairing interactions in model systems}  \and
{21.60.Cs}{Shell model}
  } % end of PACS codes
} %end of abstract
\maketitle

In many applications the infinite dimensionality of the quantum mechanical Hilbert
space is an obstacle; to overcome it, one has to restrict the model space to a finite dimensional subspace and construct an appropriate effective Hamiltonian. 
This in turn leads from a two-body to a many-body interaction terms. 
Nonetheless, the effective Hamiltonian approach has been very 
successful and even pointed to the importance of three-body nuclear interactions \cite{V3bInNP}. The recently introduced exactly solvable extended pairing model \cite{FPan04} provides a framework for study of Hamiltonians with many-body 
interaction terms:
\begin{eqnarray}
\label{Hamiltonian}
\hat{H} =\sum_{j=1}^{p}\epsilon _{j}n_{j}
&-& G\sum_{i,j=1}^{p}B_{i}^{+}B_{j}
-G \sum_{\mu =2}^{p}{\frac{1}{{{(\mu !)}^{2}}}}\times \\
&\times&\sum_{i_{1}\neq \cdots \neq i_{2\mu}}B_{i_{1}}^{+}\cdots 
B_{i_{\mu }}^{+}B_{i_{\mu +1}}\cdots B_{i_{2\mu }} \nonumber.
\end{eqnarray}

Ideally, one should be able to calculate binding energies and other observables 
{\it ab-initio} using the exact nucleon interaction. However, we are still lacking 
this capability. Instead, we use different models for binding energies and excitation energies.
Conventionally, the liquid-drop model is the zeroth order approximation to the 
binding energies while the two-body pairing interaction gives the shell model corrections. 
The extended pairing model (EPM) (\ref{Hamiltonian}) has terms beyond  the standard 
Nilsson plus pairing Hamiltonian; these terms provide an alternative description 
of  the relative binding energies (RBEs) of neighboring nuclei within the same
valence space. As we will discus below, EPM is well 
suited to provide  description of the RBEs only within the shell-model since the equations 
are insensitive to the binding energy of the core nucleus.

Beside the first two terms, Nilsson plus standard pairing intgeraction, the Hamiltonian 
in (\ref{Hamiltonian}) contains many-pair interactions which connect configurations that differ
by more than a single pair. Here $p$ is the total number
of single-particle levels considered, $\epsilon _{j}$ are single-particle
energies, $G$ is the overall pairing strength ($G>0$ ),
$n_{j}=c_{j\uparrow }^{\dagger}c_{j\uparrow} + c_{j\downarrow }^{\dagger
}c_{j\downarrow}$ is the number operator for the $j$-th single-particle
level, $B_{i}^{+}=c_{i\uparrow }^{\dagger} c_{i\downarrow}^{\dagger }$ are pair
creation operators where $c_{j}^{\dagger}$ creates a fermion in the $j$-th
single-particle level. The up and down arrows refer to time-reversed
states. Since each Nilsson level can only be occupied by one pair due to
the Pauli Exclusion Principle, the operators $B_{i}^{+}$, $B_{i}$, and
$n_{i}$ form a hard-core boson algebra: $\lbrack
B_{i},B_{j}^{+}]=\delta_{ij}(1-n_{i}),
~[B_{i}^{+},B_{j}^{+}]=0=(B_{i}^{+})^{2}$.

The pairing vacuum state $|j_{1},\cdots ,j_{m}\rangle $ is defined so that: $B_{i}|j_{1},\cdots ,j_{m}\rangle=0$ for
$1\leq i\leq p$ and $i\neq j_{s}$, where $j_{1},\cdots ,j_{m}$
indicate those $m$ levels that are occupied by unpaired nucleons.
Any state that is occupied by a single nucleon is blocked to the 
hard-core bosons due to the Pauli principle. The $k$-pair eigenstates 
of (\ref{Hamiltonian}) has the form:
\begin{equation}
|k;\zeta ;j_{1}\cdots j_{m}\rangle =\sum_{ i_{1}<\cdots <i_{k}}
C_{i_{1}\cdots i_{k}}^{(\zeta )}B_{i_{1}}^{+}\cdots
B_{i_{k}}^{+}|j_{1}\cdots j_{m}\rangle , \label{States}
\end{equation}
where $C_{i_{1}i_{2}\cdots i_{k}}^{(\zeta )}$ are expansion coefficients
 to be determined. It is assumed that the level indices
$j_{1},\cdots,j_{m}$ should be excluded from the summation in 
(\ref{States}). For simplicity, we focus only on the seniority zero case ($m=0$).

Although Hamiltonian (\ref{Hamiltonian}) contains many-body interaction
terms that are non-perturbative, the contribution of the higher and higher
energy configurations is more and more suppressed due to the structure of
the equation that needs to be solved to determine the eigensystem of the Hamiltonian
$(\ref{Hamiltonian})$. The eigensystem $E_{k}^{(\zeta )}$ and 
$C_{i_{1}i_{2}\cdots i_{k}}^{(\zeta )}$ depend on only one parameter 
$z^{(\zeta )}$, where the quantum number $\zeta$  \cite{FPan04} is understood as the $\zeta$-th solution of (\ref{z-parameter}):
\begin{eqnarray}
\label{eigenvalues}
E_{k}^{(\zeta )}&=&{z{^{(\zeta )}}}-G(k-1),\\
C_{i_{1}i_{2}\cdots i_{k}}^{(\zeta )}&=&\frac{1}{{z^{(\zeta )}}-
E_{i_{1}...i_{k}}},\quad E_{i_{1}...i_{k}}=\sum_{\mu =1}^{k}2{\epsilon }
_{i_{\mu }}, \label{Coefficints} \\
1&=&\sum_{ i_{1}<i_{2}<\cdots <i_{k}}
{\frac{G}{E_{i_{1}...i_{k}}-z{^{(\zeta )}}}}.
\label{z-parameter}
\end{eqnarray}

Due to the space limitations many details and results of the current application of 
this exactly solvable model are omitted, however, a more detailed paper is available \cite{paper2}. For the current application the single-particle energies are calculated 
using the Nilsson deformed shell model with parameters from \cite{Moller & Nix}.
Experimental BEs are taken from \cite{Audi G}. Theoretical RBE are calculated relative 
to a specific core, $^{152}$Yb, $^{100}$Sn, and $^{208}$Pb for the cases considered.
The RBE of the nucleus next to the core is used to
determine an energy scale for the Nilsson single-particle energies. For
an even number of neutrons, we considered only pairs of particles (hard
bosons). For an odd number of neutrons, we apply Pauli blocking to the
Fermi level of the last unpaired fermion and considered the remaining
fermions as if they were an even fermion system.  The valence model
space consists of the neutron single-particle levels between two closed
shells with magic numbers 50-82 and 82-126. By using (\ref{eigenvalues})
and (\ref{z-parameter}), values of $G$ are determined so that the
experimental and theoretical RBE match exactly. 

\begin{figure}[htbp]
%\resizebox{0.75\textwidth}{!}{\includegraphics{Pb-isotopes_v1c.eps}}
\centerline{\hbox{
\epsfig{figure=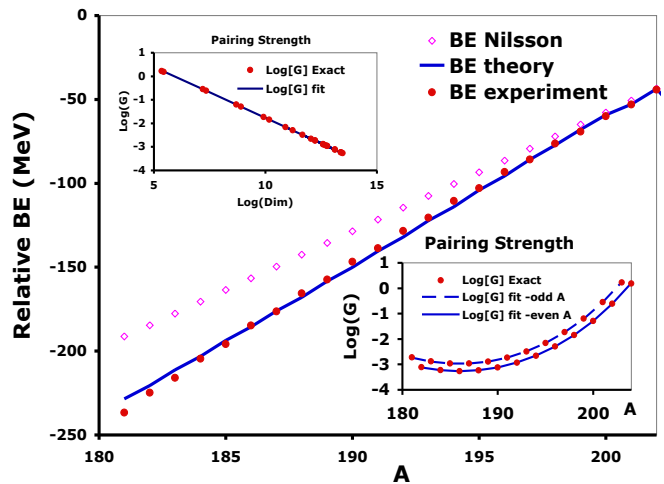,height=2.6in,width=3.5in} }}
\caption{The solid line gives the theoretical RBE for
the Pb isotopes relative to the $^{208}$Pb nucleus. The insets
show the fit to the values of $G$ that reproduce exactly the experimental
data using $^{164}$Pb core. The lower inset shows the two fitting
functions: $\log(G(A))=382.3502 - 4.1375 A + 0.0111 A^2$ for even values
of $A$ and $\log(G(A))=391.6113 - 4.2374 A + 0.0114 A^2$ for odd values
of $A$. The upper inset shows a fit to $G(A)$ that is inversely
proportional to the size of the model space, ($\dim(A)$), that is valid
for even as well as odd values of $A$: $G(A)=366.7702\dim(A)^{-0.9972}$.
The Nilsson BE energy is the lowest energy of the non-interacting system.}
\label{Pb-isotopes}
\end{figure}

Figure \ref{Pb-isotopes} shows results for the $^{181-202 }$Pb isotopes.
The RBEs are relative to $^{208}$Pb which is set to zero,
and the core nucleus is chosen to be $^{164}$Pb. For the Yb and Sn
isotopes the core nucleus is also the zero RBE reference nucleus
($^{100}$Sn and $^{152}$Yb). In this respect, the calculations for the
Pb-isotopes are different because the core nucleus ($^{164}$Pb) and the
zero binding energy reference nucleus ($^{208}$Pb) are not the same. One can
see from Figure \ref{Pb-isotopes} that a quadratic fit to $\ln(G)$ as
function of $A$ fits the data well. In this particular case, the pairing
strength $G(A)$ for all 21 nuclei in the range A=$181-202$) was also fit
to a simple two-parameter function that is inversely proportional to the
dimensionality of the model space $\dim(A)$, namely, by 
$G(A)=\alpha\dim(A)^{-\beta}$. Similar results have been obtained for the
Sn-isotopes as well using $^{132}$Sn as zero.

In conclusion, we studied RBEs of nuclei in three isotopic
chains, $^{100-130}$Sn, $^{152-181}$Yb, and $^{181-202 }$Pb, within the
recently proposed EPM \cite{FPan04} by using Nilsson
single-particle energies as the input mean-field energies. Overall, the
results suggest that the model is applicable to neighboring heavy nuclei 
and provides, within a pure shell-model approach, an alternative mean of calculating
a RBE. In order to achieve that, the pairing strength is allowed to 
change as a smooth function of the model space dimension. It is important to 
understand that the A-dependence of G is indirect, since G only depends on the model 
space dimension, which by itself is different for different nuclei. 
In particular, in all the cases studied $\ln(G)$ has a smooth quadratic behavior 
for even and odd $A$ with a minimum in the middle of the model space where the 
dimensionality of the space is a maximal; $\ln(G)$ for even $A$ and odd $A$ 
are very similar which suggests that further detailed analyses may result in the same
functional form for even $A$ and odd $A$ isotopes as found in the
case of the Pb-isotopes and Sn-isotopes. It is a non-trivial result that $G$ is inversely 
proportional to the space dimension $\dim$ in the two cases found 
(Pb-isotopes and Sn-isotopes) which requires further studies. 

\begin{acknowledgement}
Financial support provided by the U.S. National Science Foundation, the
Natural Science Foundation of China, and the Education Department of
Liaoning Province. Some of the work was also performed under the auspices
of the U. S. Department of Energy by the University of California,
Lawrence Livermore National Laboratory under contract No. W-7405-Eng-48.
\end{acknowledgement}

\end{document}